\documentclass[aps,prb,twocolumn,superscriptaddress]{revtex4}

\usepackage[dvips]{graphicx}
\usepackage[dvips]{color}
\usepackage{amsmath}

\renewcommand{\Vec}[1]{\mbox{\boldmath$#1$}}
\def\GVec#1{\mbox{\boldmath $#1$}}
\def\t{\theta}

\def\vare{\varepsilon}
\def\av#1{\langle #1 \rangle}

\newcommand{\1}{\mbox{1}\hspace{-0.25em}\mbox{l}}

\begin{document}

\title{Gate-induced Dirac cones in multilayer graphenes}

\author{Takahiro Morimoto}
\affiliation{Condensed Matter Theory Laboratory, RIKEN, Saitama, 351-0198, Japan}
\author{Mikito Koshino}
\affiliation{Department of Physics, Tohoku University, Sendai, 980-8578, Japan}

\date{\today}

\begin{abstract}
We study the electronic structures of
ABA (Bernal) stacked multilayer graphenes 
in uniform perpendicular electric field,
and show that the interplay of the trigonal warping and 
the potential asymmetry gives rise to a number of emergent 
Dirac cones nearly touching at zero energy.
The band velocity and the energy region
(typically a few tens of meV) of these gate-induced Dirac cones
are tunable with the external electric field.
In ABA trilayer graphene, in particular, 
applying an electric field induces
a non-trivial valley Hall state, 
where the energy gap at the Dirac point is filled by
chiral edge modes which propagate in opposite directions 
between two valleys. In four-layer graphene, in contrast,
the valley Hall conductivity is zero and 
there are no edge modes filling in the gap.
A nontrivial valley Hall state generally occurs
in asymmetric odd layer graphenes 
and this is closely related to a hidden chiral symmetry which 
exists only in odd layer graphenes. 
\end{abstract}

\pacs{73.22.Pr, 81.05.ue, 73.43.Cd}

\maketitle

\section{Introduction} 


Graphene is characterized with Dirac quasiparticles 
in the low-energy region, 
\cite{mcclure1956diamagnetism,divincenzo1984self,semenoff1984condensed,JPSJ.74.777}
which give rise to anomalous physical properties due to 
the linear dispersion and non-trivial Berry phase. \cite{JPSJ.67.2421,JPSJ.71.1318,PhysRevB.65.245420,PhysRevLett.95.146801,Nov05,Zha05,RevModPhys.81.109}
There are growing interests in multilayer variants of graphene such as
bilayer \cite{nov-bilayer,ohta2006controlling,ohta2007interlayer,castro2007biased}
and trilayer, 
\cite{guttinger2008coulomb,craciun2009trilayer,zhu2009carrier,bao2011stacking,jarillo-trilayer}
which also support chiral quasiparticles.
Bilayer graphene has parabolic valence
and conduction bands touching each other at Dirac point
\cite{mccann-falko,guinea-stacks06,koshino2006transport}, 
while ABA(Bernal)-stacked trilayer
graphene comes with a superposition of effective monolayer-like and
bilayer-like bands.
\cite{guinea-stacks06,latil-2006,partoens2006graphene,lu2006influence,aoki-amawashi-2007,koshino-ando07,koshino-ando08,koshino-aba-2009,koshino2009electronic,koshino-LL-2011}.
On top of these, the trigonal-warping deformation
of the energy band, which is intrinsic to graphite-based systems
\cite{slonczewski1958band,mcclure1960theory,dresselhaus-graphite02},
gives rise to small Dirac cones near Dirac point in these multilayers.
The Lifshitz transition, 
in which the Fermi circle breaks up into separate parts,
takes place at a small energy scale around a few meV.
\cite{mccann-falko,koshino2006transport,koshino-ando07,koshino-ando08,koshino-aba-2009}

On the other hand, it is possible to modify the band structure
of multilayer graphenes
by applying an electric field perpendicular to the layer,
using external gate electrodes attached to the graphene sample.
In bilayer graphene, a perpendicular electric field opens a band gap
at Dirac point.
\cite{ohta2006controlling,mccann-falko,lu2006influence,guinea-stacks06,mccann-prb-06,min-07,castro2007biased,oostinga-08}
In contrast, for ABA trilayer graphene, previous works
\cite{koshino-aba-2009,koshino-screening10}
considered the gate-field effect 
on the band structure without the trigonal warping,
and showed that the perpendicular electric field causes 
a band overlap of the conduction and valence bands at zero energy,
rather than opening a gap.
There two bands intersect each other on a circle around the K point 
whose radius is proportional to the potential asymmetry,
causing an increase of the conductivity 
at charge neutral point. \cite{craciun2009trilayer}

\begin{figure}
\begin{center}
\includegraphics[width=\linewidth]{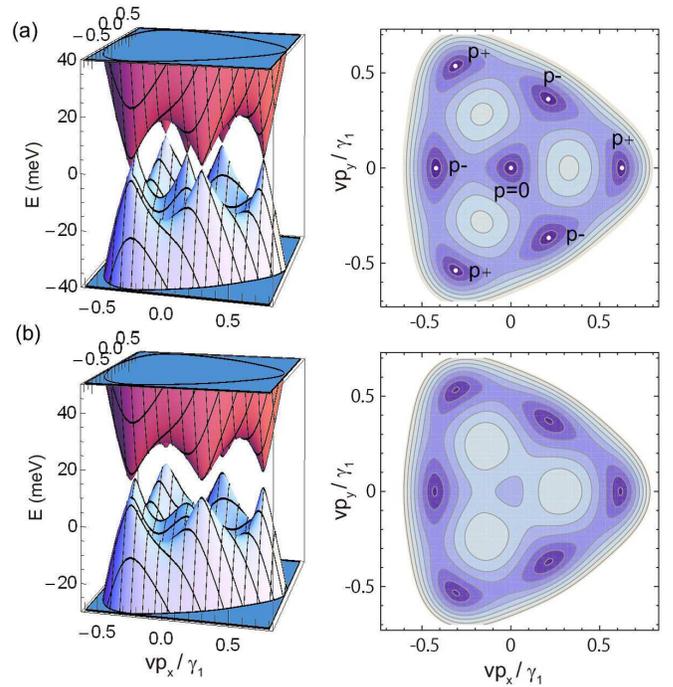}
\end{center}
\caption{
Band structures  of ABA trilayer graphene in a perpendicular 
electric field depicted in 3D plot (left panel) and contour plot (right), 
for the band models (a) only with $v,\gamma_1,v_3$ terms
and (b) with full parameters.
The interlayer potential asymmetry is set to be $\Delta=200$ meV.
}
\label{3L-band-contour}
\end{figure}

In this paper, we closely study the band structures of
ABA-stacked multilayer graphenes 
in the presence of uniform perpendicular electric field,
and find that the interplay of the trigonal warping and 
the potential asymmetry
generally gives rise to a number of additional Dirac cones
nearly touching at zero energy, as depicted
in Fig.\ \ref{3L-band-contour}(b) for trilayer graphene.
For these gate-induced Dirac cones,
the band velocity and the energy region
(i.e., the distance between Dirac point to 
the Lifshitz transition point) 
are tunable with gate bias voltage.
The energy region is typically a few tens of meV, 
which is by order of magnitude greater than in the
original non-biased multilayer graphene.
In a magnetic field, there arise triply-degenerate Landau levels 
originating from off-center gate-induced Dirac cones,
with wide energy spacings due to the linear dispersion.

The gate-induced Dirac cones are generally 
gapped at Dirac point by symmetry-breaking terms.
When the Fermi energy is in the gap, the system is in
a topologically non-trivial valley Hall state,
where electrons at $K_+$ and $K_-$ valleys 
carry opposite Hall conductivities
\cite{castro-bilayer-edge08,jung-edge-multilayer11}.
A manifestation of the valley Hall state 
is the emergence of chiral edge modes at a zigzag interface
which transports valley pseudo-spins 
in an analogous way to the spin Hall effect
\cite{murakami-qshe03,kane2005quantum}.
The valley Hall state and the helical edge modes
were previously studied for gapped monolayer and bilayer graphenes,
\cite{castro-bilayer-edge08,tse2011quantum,qiao2011electronic,qiao2011spin,qiao-gated-bilayer11,qiao-gated-bilayer-condmat12}
and also for ABC (rhombohedral) stacked trilayer graphene. 
\cite{li2012unbalanced}
We study the edge states a semi-infinite zigzag ribbon 
of asymmetric ABA multilayer graphenes,
and relate the number of the edge modes
to the valley Hall conductivity which is a bulk property.
In trilayer graphene, in particular, 
we find that non-zero valley Hall state 
is realized in a small external electric field,
and moreover, a topological transition 
takes place at a certain higher electric field, which is accompanied by
a change of the number of edge channels inside the bulk gap.
In four-layer graphene, in contrast,
the valley Hall conductivity is always zero and there are no edge
modes filling the energy gap.
We show that the nontrivial valley Hall state generally occurs
in asymmetric odd layer graphenes, and this is deeply indebted to 
an approximate chiral symmetry peculiar to odd layer graphenes.

Paper is organized as follows.
We briefly introduce the effective mass model for graphite
in Sec.\ \ref{sec_model},
and argue the trilayer graphene in Sec.\ \ref{sec_tri}
in terms of the gate-induced Dirac cones, the chiral symmetry,
the Landau level structure and and the edge states.
In Sec.\ \ref{sec_four}, we study the four-layer graphene 
as an example of even-layer cases without the chiral symmetry.
In Sec.\ \ref{sec_odd}, we argue the chiral symmetry in 
general odd-layer graphenes, generalizing the trilayer's argument.
The conclusion is given in Sec.\ \ref{sec_concl}.

\begin{figure}[tb]
\begin{center}
\includegraphics[width=0.9\linewidth]{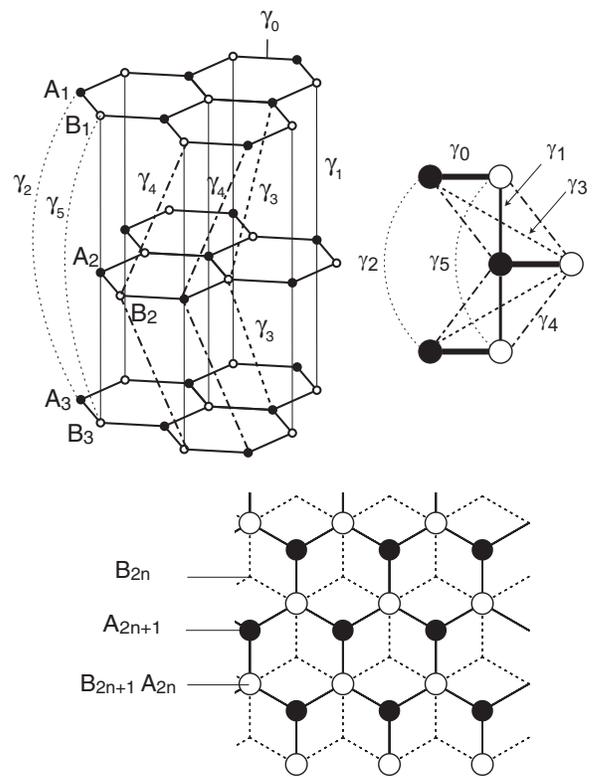}
\end{center}
\caption{
Lattice structure of ABA trilayer graphene with tight-binding hopping
parameters.
The bottom figure is a top-view, 
and the right is a schematic diagram
of the lattice structure.
}
\label{ABA-lattice}
\end{figure}

\section{Model}
\label{sec_model}

We describe the electronic properties of 
ABA-stacked multilayer graphene 
using Slonczewski-Weiss-McClure model \cite{slonczewski1958band,mcclure1960theory,dresselhaus-graphite02}
with hopping parameters described in Fig.\ref{ABA-lattice}.
The model includes intralayer coupling $\gamma_0$,
nearest interlayer couplings $\gamma_1$, $\gamma_3$ and $\gamma_4$,
next-nearest layer couplings $\gamma_2$ and $\gamma_5$,
and on-site energy asymmetry $\Delta'$,
which are estimated in bulk graphite as
\cite{dresselhaus-graphite02}:
$\gamma_0=3.16$eV, 
$\gamma_1=0.39$eV, 
$\gamma_3=0.32$eV, 
$\gamma_2=-0.020$eV, 
$\gamma_5= 0.038$eV, 
$\gamma_4= 0.044$eV,
$\Delta'=0.050$eV.
$\Delta'$ is the energy difference between the sites
which are involved in the coupling $\gamma_1$, 
and the sites which are not.

We consider ABA-stacked $N$-layer graphene,
where $|A_j\rangle$ and $|B_j\rangle$ represent
Bloch functions at the $K_{\xi}$ point, 
corresponding to the $A$ and $B$ sublattices of layer $j$,
respectively. If the basis is arranged as $|A_1\rangle,|B_1\rangle$;
$|A_2\rangle,|B_2\rangle$; $\cdots$; $|A_N\rangle,|B_N\rangle$,
the Hamiltonian in the vicinity of the $K_{\xi}$ valley is written as
\cite{guinea-stacks06,partoens2006graphene,koshino-ando07,koshino-ando08}
\begin{equation}
{\cal H}=
\begin{pmatrix}
H_0 & V & W &&\\
V^\dagger &H_0' &V^\dagger& W'&\\
W&V&H_0 & V&W\\
&W'&V^\dagger &H_0'  &V^\dagger& W'&\\
&&\ddots&\ddots&\ddots\\
\end{pmatrix}
+
\begin{pmatrix}
U_1 \\
& U_2 \\
& & U_3 \\
& & & U_4 \\
&&&&\ddots
\end{pmatrix}
\label{BernalH}
\end{equation}
with
\begin{eqnarray*}
&& H_0=
\begin{pmatrix}
0 & v \pi^\dagger \\
v \pi & \Delta'  \\
\end{pmatrix},
H_0'=
\begin{pmatrix}
\Delta' & v \pi^\dagger \\
v \pi & 0  \\
\end{pmatrix},\\
&&
V=
\begin{pmatrix}
-v_4 \pi^\dagger & v_3 \pi \\
\gamma_1 & -v_4 \pi^\dagger \\
\end{pmatrix},
\\
&& W=
\begin{pmatrix}
\gamma_2/2 & 0 \\
0 & \gamma_5/2 \\
\end{pmatrix},
W'=
\begin{pmatrix}
\gamma_5/2 & 0 \\
0 & \gamma_2/2 \\
\end{pmatrix},
\\
&&U_{j} = 
-\Delta
\left(j-\frac{N+1}{2}\right)
\begin{pmatrix}
1 & 0 \\
0 & 1 \\
\end{pmatrix}.
\end{eqnarray*}
where $\pi=\xi \pi_x+i \pi_y$, 
$\Vec{\pi} = \Vec{p}+e\Vec{A}$ with 
$\Vec{A}$ being the vector potential arising from the 
applied magnetic field,
and $\xi=\pm 1$ are the valley indeces for $K_\pm$.
The parameter $v=\sqrt{3} a \gamma_0/(2 \hbar)$ is the 
band velocity for monolayer graphene, and
$v_3=\sqrt{3} a \gamma_3/(2 \hbar)$ a velocity 
related to the band parameter $\gamma_3$,
where $a \approx 0.246$ nm is the distance 
between the nearest $A$ sites on the same layer.
$U_j$ describes the electrostatic potential on $j$-th layer 
induced by the external electric field,
where we assumed a uniform potential gradient 
in the perpendicular direction,
This is valid in a few-layer graphene with typically $N <\sim 5$.
For thicker multilayers with $N >\sim 10$,
the potential drop occurs within a few layers near the external gate
due to the screening by the charge carriers in graphene.
\cite{koshino-aba-2009,koshino-screening10}


\section{Trilayer graphene}
\label{sec_tri}

\subsection{Chiral symmetry and gate-induced Dirac cones}

The Hamiltonian of ABA-trilayer graphene is given by
Eq.\ (\ref{BernalH}) with $N=3$,
where the external potential is 
$(U_1,U_2,U_3) = (\Delta,0,-\Delta)$.
In the absence of $\Delta$, 
the Hamiltonian can be block diagonalized into monolayer-like band
and bilayer-like band. \cite{koshino-aba-2009}
In finite $\Delta$, these sub-blocks are hybridized,
and it is then useful to arrange the basis as 
\begin{eqnarray}
&&  
\bigl\{(|A_1\rangle-|A_3\rangle)/\sqrt 2,
\,\, |B_2\rangle,
\,\, (|B_1\rangle+|B_3\rangle)/\sqrt 2, 
\nonumber \\
&&  \quad
(|B_1\rangle-|B_3\rangle)/\sqrt 2,
\,\,|A_2\rangle,
\,\,(|A_1\rangle+|A_3\rangle)/\sqrt 2\bigr\},
\label{eq_base}
\end{eqnarray}
where the monolayer-like band correspond to 1st and 4th bases  
and bilayer-like band to 2nd, 3rd, 5th, and 6th.
We categorize the first three bases in Eq.\ (\ref{eq_base}) 
as group $\circ$, and the last three as group $\bullet$.
The Hamiltonian is written  in this basis as
\begin{eqnarray}
&& {\cal H}=
\begin{pmatrix}
H_{\circ} & D_- \\
D_+ & H_{\bullet} 
\end{pmatrix}
\label{H-tri-f}
\\
&&
D_+
= \left(
\begin{array}{ccc}
v\pi & 0 & \Delta \\
0 & v\pi^\dagger & \sqrt 2 \gamma_1 \\
\Delta & \sqrt 2 v_3 \pi & v \pi^\dagger 
\end{array}
\right), \quad D_- = (D_+)^\dagger
\\
&&
H_{\circ}=
\left(
\begin{array}{ccc}
-\gamma_2/2 & 0 & 0 \\
0 & 0 &  -\sqrt{2}v_4\pi \\
0 & -\sqrt{2}v_4\pi^\dagger & \gamma_5/2+\Delta' 
\end{array}
\right)
\\
&&
H_{\bullet}=
\left(
\begin{array}{ccc}
-\gamma_5/2+\Delta' & 0 & 0 \\
0 & \Delta' &  -\sqrt{2}v_4\pi \\
0 & -\sqrt{2}v_4\pi^\dagger & \gamma_2/2 
\end{array}
\right)
\end{eqnarray} 

If we keep only relevant band parameters $\gamma_0$,
$\gamma_1$, $\gamma_3$ and the potential $\Delta$,
and neglect remaining parameters, 
the Hamiltonian Eq.\ (\ref{H-tri-f}) 
possesses the chiral symmetry (sublattice symmetry),
in that the diagonal matrix blocks $H_{\circ}$ and $H_{\bullet}$
all vanish, leaving the off-diagonal blocks $D_\pm$ 
which connects the bases of $\circ$ to the bases of $\bullet$.
The reason for this can be understood in terms of reflection symmetry
as follows: In the group $\circ$($\bullet$), the bases 
associated with $A$ and $B$ sublattices
have odd and even (even and odd)
parity, respectively, with respect to the reflection in the middle layer.
The Hamiltonian without $\Delta$,
i.e., the first term in Eq.\ (\ref{BernalH})
has even parity in the reflection,
and has matrix elements only between $A$ and $B$ sublattices 
when only $\gamma_0$, $\gamma_1$ and $\gamma_3$ are kept.
After the unitary transformation,
it does not give any matrix elements in $H_\circ$ or $H_\bullet$
because in each group ($\circ$ or $\bullet$),
a base associated with $A$ and one associated with $B$
always have different parity from the definition.
On the other hand, the potential term, i.e.,
the second term in Eq.\ (\ref{BernalH}) is odd in the reflection
and matrix elements only connect the same sublattice.
It gives no matrix elements in $H_\circ$ or $H_\bullet$ either,
because in each group, bases associated with the same sublattice 
always have the same parity.

The energy spectrum of this simplified Hamiltonian
contains a single center Dirac cone and six off-center Dirac cones 
at zero energy, as depicted in Fig.\ref{3L-band-contour}(a). 
The robustness of gapless spectrum is protected by the chiral symmetry.
The extra terms with $\gamma_2, \gamma_5, v_4$ and $\Delta'$
in the diagonal blocks break the chiral symmetry and open small energy gaps 
at these Dirac cones as in Fig.\ref{3L-band-contour}(b).

The positions of the Dirac points 
in the chiral Hamiltonian without $H_{\circ}$ and $H_{\bullet}$
can be found by solving $ {\rm det} \, D_+ =0$
with $\pi = \xi p_x + i p_y= p e^{i \theta}$. 
We obtain a Dirac point at $p=0$,  
and six off-center Dirac points at
\begin{eqnarray}
&&p=p_+; \quad \theta= 0, \frac{2 \pi}{3},\frac{4 \pi}{3}, 
\nonumber\\
&&p=p_-; \quad  \theta= \frac{\pi}{3}, \pi, \frac{5 \pi}{3}, 
\nonumber\\
&& p_\pm= \frac{\pm \gamma_1 v_3 + \sqrt{\Delta^2 v^2+\gamma_1^2 v_3^2}}{v^2}.
\end{eqnarray}

\begin{figure*}[tb]
\begin{center}
\includegraphics[width=0.95\linewidth]{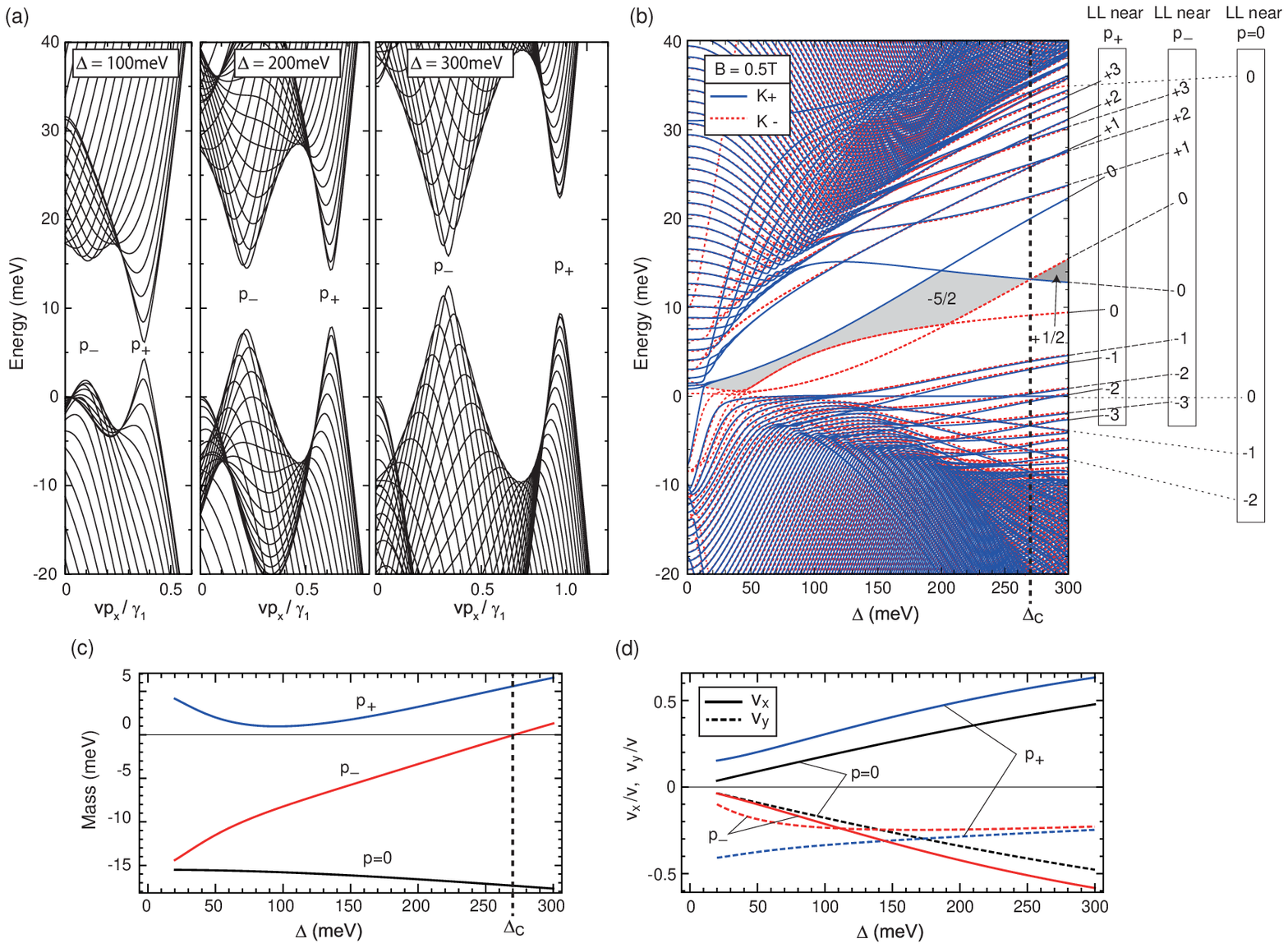}
\end{center}
\caption{
(a) Band structure of gated ABA trilayer with various potential
 asymmetry $\Delta$.
(b)Gate bias dependence of Landau level energy in $B=0.5$T.
Landau levels in $K_+$ and $K_-$ valleys are 
plotted as solid (blue) and dashed (red) lines, respectively. 
Shaded area represents the charge neutral region,
where the number indicates the single valley Hall conductivity
$\sigma_{xy}^{K\xi}$ in units of $\xi e^2/h$
(c) Effective mass  $m$ and (d) band velocities $v_x,v_y$ 
for the effective Dirac cones at $p=0,p_+ ,p_-$,
plotted against $\Delta$.
}
\label{3L-LL}
\end{figure*}

At each Dirac point, the degenerate zero-energy bases 
$\psi_\circ$ and $\psi_\bullet$ are derived
from the equation
$D_+  |\psi_\circ\rangle = 0$ and $D_- |\psi_\bullet\rangle = 0$,
and the effective Dirac Hamiltonian is given by
\begin{eqnarray}
 H_{\rm eff} =
\begin{pmatrix}
0 & \langle \psi_\circ| D_- |\psi_\bullet\rangle \\
\langle \psi_\bullet| D_+ |\psi_\circ\rangle & 0
\end{pmatrix},
\end{eqnarray}
keeping the lowest order in the momentum shift from the Dirac point.
By rotating $(x,y)$ coordinate
and the spinor space by angle $\theta$
at the same time, this is transformed to
\begin{eqnarray}
 H_{\rm eff} = v_x \tilde\pi_x \sigma_x + v_y \tilde\pi_y \sigma_y,
\label{analytical-Heff}
\end{eqnarray}
where $\sigma_x$ and $\sigma_y$ are Pauli matrices,
and $(\tilde\pi_x,\tilde\pi_y)$ is the momentum measured from each Dirac point.
We find for the Dirac point at $p=0$ (no need for rotation), 
\begin{eqnarray}
&& \psi_\circ =(0,1,0) \nonumber\\
&& \psi_\bullet=(-\sqrt{2}\gamma_1,\Delta,0)/\sqrt{\Delta^2+2\gamma_1^2}, 
\end{eqnarray}
and
\begin{eqnarray}
&& v_x = \xi \frac{\Delta v}{\sqrt{\Delta^2+2\gamma_1^2}},
\quad
v_y = -\frac{\Delta v}{\sqrt{\Delta^2+2\gamma_1^2}}.
\end{eqnarray}
with the valley index $\xi = \pm$.
For the off-center Dirac points at $p =p_\pm$,
\begin{eqnarray}
&& \psi_\circ=
(-\Delta e^{i\theta}, \mp \sqrt{2}\gamma_1, vp_\pm e^{2i\theta})/C_1, 
\nonumber\\
&& \psi_\bullet =
(-\Delta, \mp \sqrt{2}v_3p_\pm,  vp_\pm e^{-i\theta})/C_2, \nonumber\\
&& C_1= \sqrt{\Delta^2+2\gamma_1^2+v^2p_\pm^2}, 
 C_2= \sqrt{\Delta^2+(2v_3^2+v^2)p_\pm^2},
\nonumber\\
\end{eqnarray}
and the velocities after rotation of $(x,y)$ by $\theta$ become
\begin{eqnarray}
&& v_x =
2\xi v(
  \pm \Delta^2  + \gamma_1 v_3  p_\pm)/(C_1C_2)
\nonumber\\
&& v_y =
- 6 \gamma_1 v_3  v p_\pm/(C_1C_2),
\label{eq_v_off_center}
\end{eqnarray}
where $v_x$ and $v_y$ correspond to the radial and azimuthal 
directions respectively, with respect to $p=0$.
$v_x$ and $v_y$ 
of each Dirac points are plotted in Fig.\ \ref{3L-LL}(d).
The velocities are mainly enhanced
by applying $\Delta$, while $v_y$ for $p_+$ decreases only slowly.
This indicates that the conductivity, which is roughly proportional to 
the square of the band velocity \cite{JPSJ.67.2421},
is enhanced by $\Delta$, as is consistent with the previous 
transport measurement 
\cite{craciun2009trilayer}
and the theoretical estimation \cite{koshino-aba-2009}.


The chirality for each Dirac cone can be defined by
$\nu_c \equiv \textrm{sgn}(v_x v_y)$,
and this coincides with the Berry phase in units of $\pi$ 
around the Dirac point.
We find $\nu_c = -\xi$ for $p=0$, and $\nu_c = \mp \xi$ for $p=p_\pm$,
so that the summation of chirality over seven Dirac points
in the valley $K_\xi$ is $-\xi$.
Since the chirality is a topologically protected number 
as long as the chiral symmetry is present,
non-zero total chirality indicates that 
the conduction band and the valence band 
inevitably touch at some points in any value of $\Delta$.


Now we consider all the band parameters in Eq. (\ref{H-tri-f})
to argue the energy gaps at the Dirac points.
The effective Hamiltonian of each gate-induced Dirac cone is modified to
\begin{eqnarray}
 H_{\rm eff}=v_x \tilde \pi_x \sigma_x + v_y \tilde \pi_y \sigma_y 
+ m \sigma_z + \epsilon_0 I,
\label{eq_massive_dirac}
\end{eqnarray}
where the mass $m$ and the energy shift $\epsilon_0$ are given by
\begin{eqnarray}
&& m= [\langle \psi_\circ| H_\circ  |\psi_\circ\rangle
-\langle \psi_\bullet| H_\bullet  |\psi_\bullet\rangle]/2,
\\
&& \epsilon_0 = [\langle \psi_\circ| H_\circ  |\psi_\circ\rangle
+\langle \psi_\bullet| H_\bullet  |\psi_\bullet\rangle]/2.
\end{eqnarray}
The width of the energy gap is $2m$.
Note that the approximation using the gapped Dirac Hamiltonian above
is valid when the mass gap 
is smaller than the energy region of the gate-induced Dirac cone
below the Lifshitz transition point.
In Fig.\ref{3L-LL}(c),
we show the evaluated mass $m$ for the seven Dirac points.
We see that the mass for $p=p_-$ changes from negative to positive
when $\Delta$ exceeds the critical value,
\begin{eqnarray}
 \Delta_c \approx 270\, {\rm meV},
\end{eqnarray}
at which the energy gap closes.

\begin{table}
 \[
  \begin{array}{c|ccc}
p & 0(\times 1) & p_+ (\times 3)  & p_- (\times 3)  
\\
\hline
{\rm sgn}(v_xv_y) & -\xi  & -\xi & +\xi
\\
\hline
{\rm sgn}(m) & -1  & +1 & 
\begin{cases}
-1 ~~ (\Delta<\Delta_c) \\
+1 ~~ (\Delta>\Delta_c)
\end{cases}
\\
\hline
{\rm sgn}(v_xv_ym) & +\xi  & -\xi & 
\begin{cases}
-\xi ~~ (\Delta<\Delta_c) \\
+\xi ~~ (\Delta>\Delta_c)
\end{cases}
\end{array}
 \]
\caption{Sign of chirality, mass and in-gap Hall conductivity
for gate-induced Dirac cones at $K_\xi$ valley.}
\label{tbl_chirality}
\end{table}


When the Fermi energy lies in the gap 
in massive Dirac Hamiltonian Eq.\ (\ref{eq_massive_dirac}), 
the Hall conductivity takes non-zero value
$\sigma_{xy} = [e^2/(2 h)]{\rm sgn}(v_x v_y m)$ 
even in zero magnetic field. \cite{haldane-qhe,oshikawa94}
Considering chirality and mass 
for each Dirac point in Table \ref{tbl_chirality},
the total Hall conductivity summed over
the Dirac points at single valley $K_\xi$ becomes
\begin{eqnarray}
\sigma_{xy}^{K_\xi} =
\frac{e^2}{h} \times
\left\{
\begin{array}{cc}
\displaystyle -\frac{5}{2}\xi & (\Delta<\Delta_c), 
\vspace{2mm}\\
\displaystyle +\frac{1}{2}\xi   & (\Delta>\Delta_c).
\end{array}
\right.
\end{eqnarray}
We have a  topological change at gap closing point, $\Delta = \Delta_c$.
The Hall conductivity have opposite signs between two valleys $K_\pm$
due to the time-reversal symmetry, so that the net Hall 
conductivity is always zero.
Nevertheless, the single-valley Hall conductivity is directly related to
the number of chiral edge modes appearing in zigzag edge, 
as we will see in Sec. \ref{sec:edge-modes}.


\subsection{Landau level structure}

The Landau levels in the presence of a uniform magnetic field 
$\Vec{B}=(0,0,B)$ can be calculated by the Hamiltonian with
$\pi = (\sqrt{2}\hbar/\ell) a^\dagger$ and $(\sqrt{2}\hbar/\ell) a$
for $K_+$ and $K_-$, respectively.
Here $a^\dagger$ and $a$ are raising and lowering operators, respectively,
which operate on the Landau-level wave function $\phi_n$ as
$a \phi_n = \sqrt{n}\phi_{n-1}, 
a^\dagger \phi_n = \sqrt{n+1}\phi_{n+1}$, 
and $\ell = \sqrt{\hbar/(eB)}$ is the magnetic length.
The Landau level spectrum at $B=0.5$T
is plotted as a function of 
potential asymmetry $\Delta$ in Fig.\ref{3L-LL}(b).
Landau levels in $K_+$ and $K_-$ valleys are 
plotted as solid (blue) and dashed (red) lines, respectively.
There we can see two distinct regions,
a region around zero energy where the Landau level spacing are wide due 
to Dirac Landau levels,
and a region above the Lifshitz transition
where the Landau levels are densely spaced
due to large density of states.
In increasing $\Delta$, 
the gate-induced Dirac pockets accommodate
more and more Landau levels
as the energy region within the Dirac pockets expands.

The low-energy Landau levels inside the gate-induced Dirac pockets
can be approximately described by
the massive Dirac Hamiltonian, Eq.\ (\ref{eq_massive_dirac}).
There the Landau level spectrum is explicitly written as 
\begin{eqnarray}
\begin{cases}
E_{0} = \epsilon_0 - {\rm sgn}(v_xv_y) \, m,
\\
E_{\pm n} = \epsilon_0 \pm \sqrt{\Delta_B^2 n + m^2},~~~(n>0),
\end{cases}
\end{eqnarray}
where
\begin{eqnarray}
 \Delta_B = \sqrt{2 |v_xv_y| \hbar e B}.
\end{eqnarray}
We note that the Landau level of $n=0$ is sensitive 
to the chirality and the sign
of the mass: it appears at the top of the valence band and the 
bottom of the conduction band when
${\rm sgn}(v_xv_y m) >0$ and $<0$, respectively.
The two cases correspond to the mid-gap values of the Hall conductivity,
$\sigma_{xy}= e^2/(2h)$ and $-e^2/(2h)$, respectively.
Also, since the chirality ${\rm sgn}(v_xv_y)$ is opposite
between $K_+$ and $K_-$ valleys,
the $n=0$ Landau level split in valleys, while all others are valley 
degenerate. Besides, we have additional triple-fold 
degeneracies for each of $p_+$ and $p_-$.

In Fig.\ref{3L-LL}(b), 
we actually see that the levels 
$n=0$ level is non-degenerate 
in valleys, appearing at either of the band edges
of each massive Dirac band.
At $\Delta = \Delta_c$, we observe that
one pair of $n=0$ levels from $K_\pm$ valleys crosses
each other at the charge neutral point,
in accordance with the topological change 
of $\sigma_{xy}^{K_\xi}$ from $(-5/2)\xi e^2/h$ to $ (+1/2)\xi e^2/h$.
The levels $n>0$ are almost valley-degenerate while 
tiny splitting is due to the deviation from the massive Dirac Hamiltonian
Eq.\ (\ref{eq_massive_dirac}).

When we drop the band parameters other than 
$\gamma_0,\gamma_1$ and $\gamma_3$, 
the Hamiltonian becomes chiral symmetric and 
the zero-th Landau level $E_0$ of each Dirac cone
comes exactly to zero energy as a chiral zero mode,
of which wavefunction has amplitude only on
$\circ$ and $\bullet$ for $\nu_c = \mp \xi$, respectively.
The index theorem then states that the difference between 
the number of the zero-modes belonging to 
$\circ$ $(n_+)$  and those to $\bullet$ $(n_-)$,
is defined as chiral index, 
which coincides with the gauge flux penetrating the system.
The chiral index in the present case is shown to be 
$n_+ - n_- = \xi \Phi$ where $\Phi = eBS/h$
is the magnetic flux penetrating the system area $S$.
This is, in units of $\Phi$, 
coherent with a summation of $-\nu_c$ in each single valley.
 As in the conventional Dirac Hamiltonian, 
\cite{jackiw-rebbi-index77,fujikawa79,fujikawa-anomaly}
the chiral index can be related to
the geometric curvature of the gauge field,
and the above relation between the chiral index and total magnetic flux
stands in non-uniform magnetic field as well.
The detailed argument is presented in Appendix.\ \ref{sec:chiral}.




\subsection{Edge modes} \label{sec:edge-modes}

\begin{figure}
\begin{center}
\includegraphics[width=0.8\linewidth]{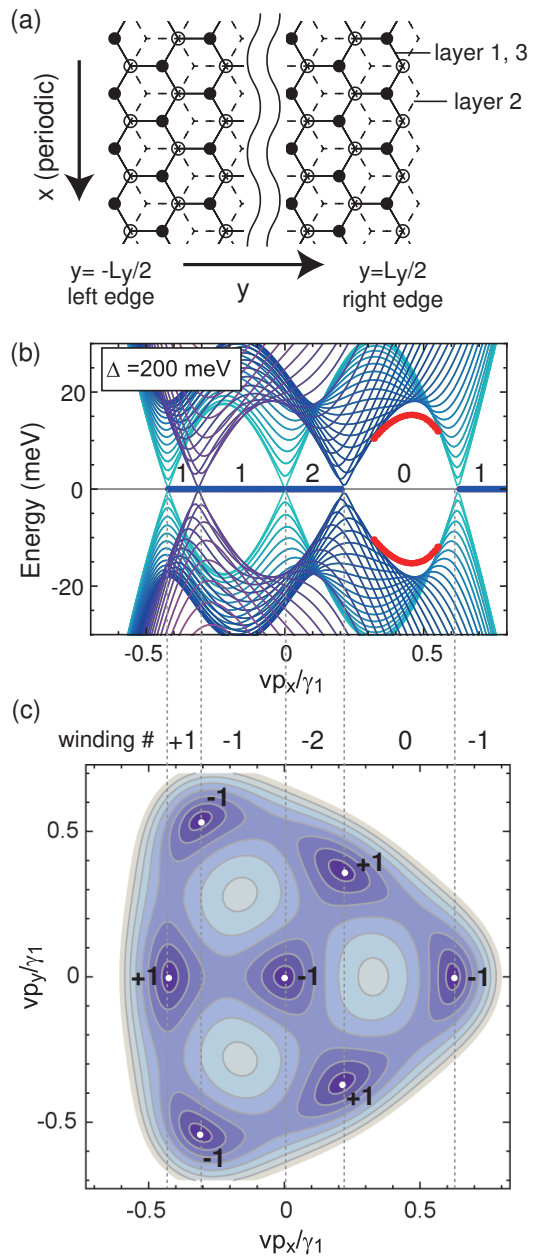}
\end{center}
\caption{
(a) Atomic structure of a zigzag ribbon of ABA trilayer graphene.
(b) Band structure around $K_+$ valley
of a zigzag ribbon of ABA trilayer graphene 
only including $\gamma_0, \gamma_1$ and $\gamma_3$,
with the asymmetric potential $\Delta=200 $ meV.
Bulk, left-edge and right-edge states are plotted with 
solid lines, red dots and blue dots, respectively.
Numbers indicate the degeneracies of 
the edge state bands for a single side
(the same for the left and the right edges).
(c) Contour plot of the bulk band structure 
with white dots and labels denoting the position and 
the chirality of gate-induced Dirac points, respectively.
One-dimensional winding number $\gamma(k_x)$ 
is indicated between dotted lines penetrating the Dirac points.
}
\label{3L-zigzag-chirality}
\end{figure}

Non-trivial Hall conductivity in single valley
indicates an existence of chiral edge modes localized at the interface,
as long as the valley mixing is not present.
There the number of emergent edge modes are directly related 
to the Hall conductivity,
so that chiral edge modes as many as the number 
of $\sigma_{xy}^{K_\pm}$ should counterflow in opposite directions between 
$K_+$ and $K_-$,
as is analogous to the spin Hall insulator.\cite{murakami-qshe03}
Here we numerically examined the edge modes in 
the asymmetric ABA trilayer graphene with zigzag
interfaces, in which the valley mixing is absent.
We consider a semi-infinite system with a zigzag boundary
along $x$ direction as shown in Fig. \ref{3L-zigzag-chirality}(a), 
where $p_x$ is a good quantum
number.  The energy of edge modes in the bulk gap can be obtained by
searching for the evanescent modes satisfying a boundary condition
at the interface. The method is detailed in Appendix \ref{sec:edge}.

\begin{figure}
\begin{center}
\includegraphics[width=0.8\linewidth]{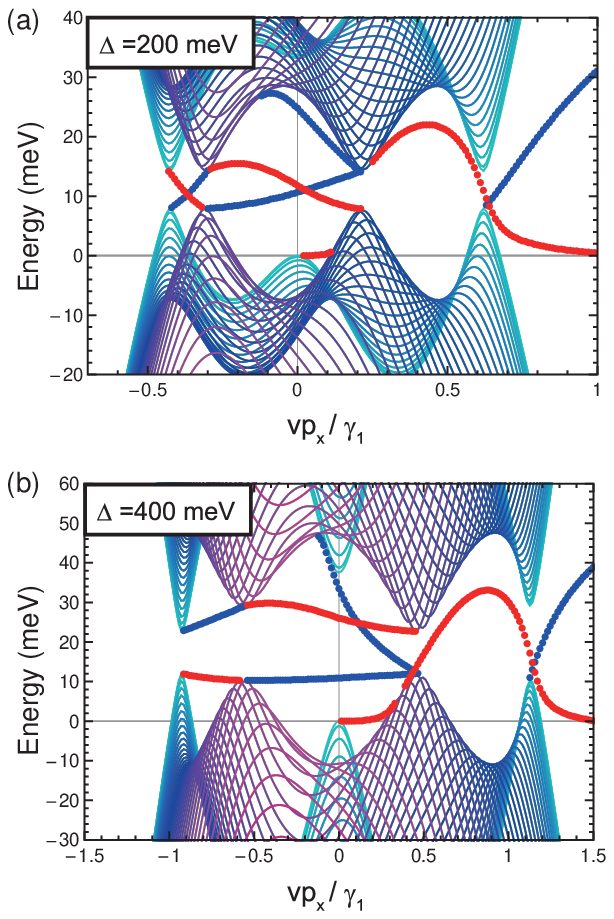}
\end{center}
\caption{
Band structure around $K_+$ valley
of zigzag ribbon of ABA trilayer graphene with the parameters
fully included, in the asymmetric potential 
(a) $\Delta=200 $ meV and (b)  $\Delta=400 $ meV.
Bulk, left-edge and right-edge states are plotted with 
solid lines, red dots and blue dots, respectively.
}
\label{3L-zigzag-full}
\end{figure}

First we consider the chiral symmetric case neglecting 
$\gamma_2, \gamma_5, v_4, \Delta'$.
Fig.\ref{3L-zigzag-chirality} illustrates 
the energy spectrum near $K_+$ point at $\Delta=200$ meV,
where we see that zero-energy edge modes 
appear between some of gate-induced Dirac points.
The number of zero energy edge
modes are closely related to the chirality of each Dirac cone.
\cite{ryu_hatsugai_2002,schnyder-ryu11,heikkila2011dimensional,burkov2011topological}
When we regard a two-dimensional periodic system on $xy$-plane
as a one-dimensional system with fixed $p_x$ as a parameter,
we can define a winding number $\gamma(p_x)$
by integrating the Berry phase change
all the way along $p_y$  on the Brillouin zone at the fixed $p_x$. 
When we set the boundary perpendicular to $y$ axis
(i.e., $p_x$ is still a good quantum number),
the number of zero-energy edge modes appearing at the boundary
coincides with $\gamma(p_x)$ except a constant.\cite{ryu_hatsugai_2002}
In the present system, this bulk-edge relationship can be clearly seen in
Fig.\ref{3L-zigzag-chirality}.

Other hopping terms breaking the chiral symmetry
give rise to mass to the Dirac points,
and relative signs between these masses
determines the connection of the chiral edge modes
between different Dirac points.
In Fig.\ref{3L-zigzag-full}(a), we plot the band structure 
near $K_+$ including full band parameters at $\Delta=200 $ meV. 
We see that the left and right edge modes stick to either of 
the top or bottom of gapped Dirac cone depending on the sign of the
mass.  At the charge neutral point, 
we have three set of chiral edge channels crossing the Fermi energy,
which all circulate in clockwise direction when viewed from $+z$ direction.
We also observe two edge states extending out of the plot
and leading to the other valley $K_-$.
In $K_-$, we have the exactly same spectrum 
with $p_x$ inverted to $-p_x$.

The correspondence to the single-valley Hall conductivity
can be understood in a similar argument
to that for integer quantum Hall effect.\cite{laughlin1981qhc}
Let us consider a cylindrical system which is
closed in $x$ with circumference $L_x$ 
while finite in the axial direction $y$ with $-L_y/2 < y < L_y/2$
bound by the zigzag edges. 
When we adiabatically turn on 
a magnetic flux quantum $h/e$ penetrating into the cylinder
(inducing an electric field along $-x$ direction), 
every state at $p_x$ is shifted to $p_x+2\pi\hbar/L_x$.  
The single-valley Hall conductivity of $K_+$ 
then coincides with the total move of $K_+$ electrons 
in $-y$ direction through this adiabatic process.
At the Fermi energy, an electron moves from the 
left edge to the right for each of three pairs 
of counter-propagating channels, 
contributing to $\sigma_{xy}^{K_+} = -3 e^2/h$.
A charge transfer also occurs below the Fermi energy,  
where a bulk state $(\av y=0)$
is pumped to an edge state $(\av y= -L_y/2)$
in the left-edge channel going out of the valley.
This yields a contribution to $\sigma_{xy}^{K_+} = (+1/2) e^2/h$,
which adds up to $\sigma_{xy}^{K_+} = (-5/2) e^2/h$ all together.

Fig.\ref{3L-zigzag-full}(b) plots the band structure at
larger bias,
$\Delta=400 $ meV after the topological transition at $\Delta_c$.
We find that the connection of the edge modes changes
at $p=p_-$, leaving only one clockwise and one anti-clockwise
chiral edge modes crossing at the Fermi energy.
The Hall conductivity from those two exactly cancel out, 
while we have the same contribution 
from the edge channel below the Fermi energy,
giving the total Hall conductivity $\sigma_{xy}^{K_+} = (+1/2) e^2/h$.
This again coincides with bulk valley Hall conductivity 
estimated from the mass and chirality.
Conversely, the single valley Hall conductivity
gives the number of the counter edge modes
crossing at the Fermi energy,
when we appropriately exclude the half integer contribution 
from the edge modes connecting $K_\pm$.

\section{Four-layer graphene}
\label{sec_four}

Unlike trilayer, a four-layer graphene with interlayer
asymmetry does not possess the chiral symmetry even
in the approximate model, and the band gap always open at the zero energy. 
The effective low-energy Hamiltonian can be obtained 
by excluding the high-energy bonding states at $|\vare| \sim O(\gamma_1)$
as $H_{\rm eff} = H_{11} - H_{12}H_{22}^{-1}H_{21}$,\cite{mccann-falko}
where $H_{11}$ and $H_{22}$ represent diagonal blocks 
of the original Hamiltonian
for low-energy bases spanned by $(A1,B2,A3,B4)$,
and for high-energy bases by $(B1,A2,B3,A4)$, respectively,
and $H_{12}$ and $H_{21}$ are off-diagonal blocks connecting them.
This is explictly written in basis $(A1,B2,A3,B4)$ as
\begin{eqnarray}
H_{\rm eff} &=&
\frac{v^2}{\gamma_1}
\begin{pmatrix}
0 &  -(\pi^\dagger)^2 & 0 &  (\pi^\dagger)^2 \\
-\pi^2 & 0 & 0& 0 \\
 0 &0 & 0 &  -(\pi^\dagger)^2  \\
 \pi^2 & 0 &   -\pi^2 &  0 \\
\end{pmatrix}
\nonumber \\
&+&
v_3
\begin{pmatrix}
0& \pi &&\\
\pi^\dagger&0&\pi^\dagger& \\
&\pi&0&\pi \\
&&\pi^\dagger&0      \\
\end{pmatrix}
+
\Delta
\begin{pmatrix}
\frac{3}{2}\\
& \frac{1}{2}\\
& & -\frac{1}{2}\\
& & & -\frac{3}{2}
\end{pmatrix},
\end{eqnarray}
where we neglected the band parameters other than
$\gamma_0, \gamma_1$ and $\gamma_3$.
The approximation is valid only when $vp, \Delta \ll \gamma_1$.

If we even neglect $v_3$ term,
the low-energy energy band is rotationally symmetric around $K\pm$ points 
and its dispersion relation is given by
\begin{eqnarray}
\epsilon(p)= 
\pm 
\frac{1}{2} 
\sqrt{6 \tilde \epsilon^2 +5 \Delta^2
-2 \sqrt{5 \tilde \epsilon^4 
         +20 \tilde \epsilon^2  \Delta^2+4 \Delta^4}}
\end{eqnarray}
with $\tilde \epsilon=v^2 p^2/\gamma_1$.
The band gap appears 
between $\vare = \pm (\Delta/2)(-7+16/\sqrt{5})^{1/2}$,
corresponding to an off-center momentum
$p_0 = (-2+6/\sqrt{5})^{1/4}\sqrt{\gamma_1\Delta}/v$.

\begin{figure}[tb]
\begin{center}
\includegraphics[width=\linewidth]{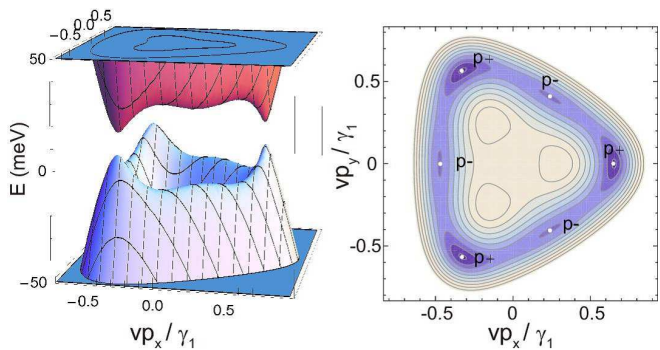}
\end{center}
\caption{
Band structures of ABA-stacked fourlayer graphene
with the interlayer asymmetric potential $\Delta=100$ meV,
depicted in 3D plot (left panel) and contour plot (right).
}
\label{4L-band-contour}
\end{figure}

\begin{figure*}[tb]
\begin{center}
\includegraphics[width=0.85\linewidth]{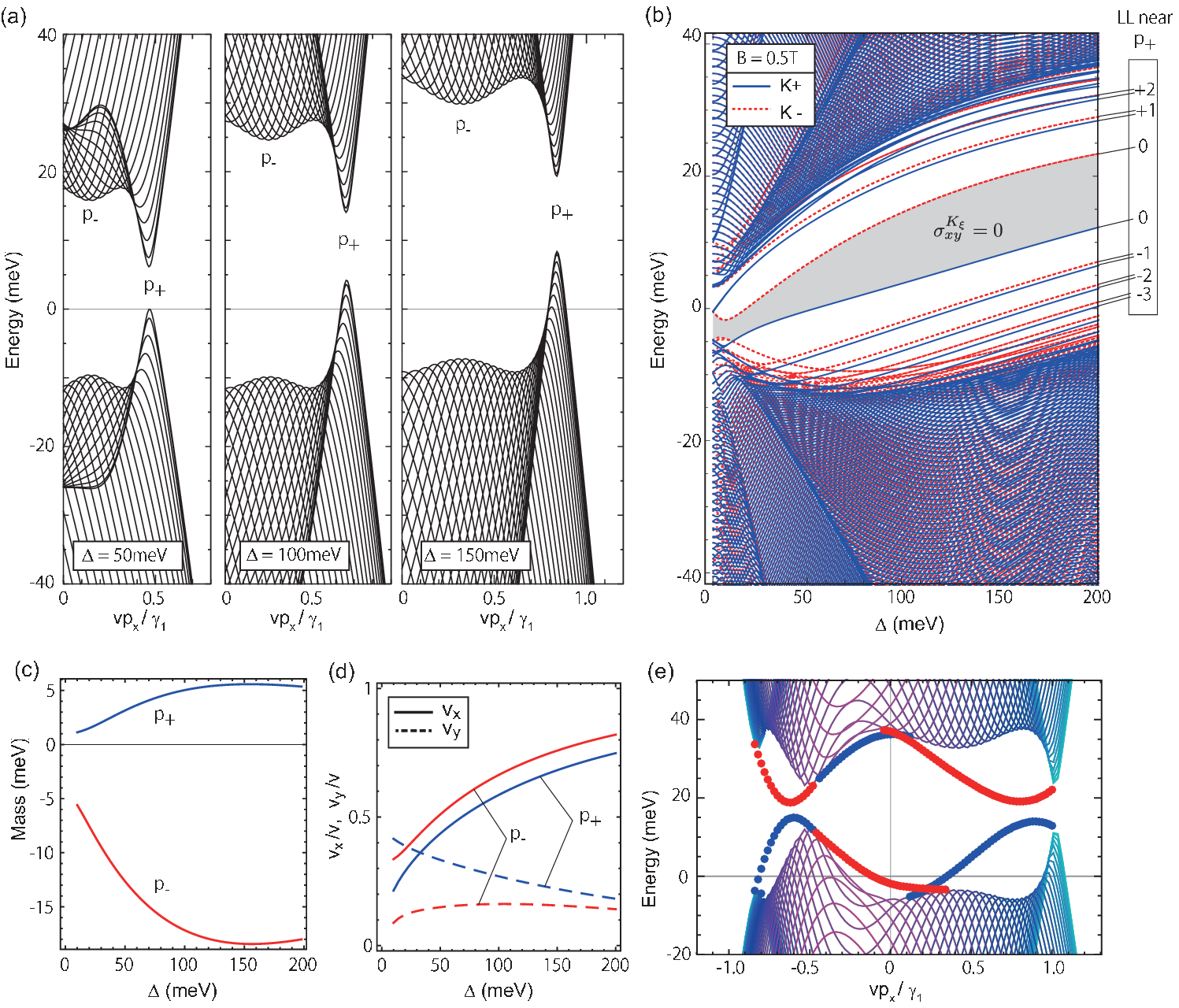}
\end{center}
\caption{
Plots similar to Fig.\ \ref{3L-LL} for ABA fourlayer graphene.
(a) Band structure with various potential asymmetry $\Delta$.
(b) Gate bias dependence of Landau level energy in $B=0.5$T.
Shaded area represents the charge neutral region.
(c) Effective mass  $m$ and (d) band velocities $v_x,v_y$ 
for the effective Dirac cones at $p_+ ,p_-$,
plotted against $\Delta$.
(e) Band structure similar to Fig.\ \ref{3L-zigzag-full}
for zigzag ribbon of ABA fourlayer with $\Delta=200$meV.
}
\label{4L-LL}
\end{figure*}

When we resume $v_3$ term and other parameters,
six off-center pockets emerge at momentum $p_+$ and $p_-$ near $p_0$, 
each of which are arranged in 120 degrees symmetry
as illustrated in Fig.\ref{4L-band-contour},
and Fig.\ref{4L-LL}(a).
The pre-existent gap never closes during this process.
The pockets at $p=p_+$ are much deeper than those at $p=p_-$,
and the energy depth is about 15 meV at $\Delta=100$ meV.
Fig.\ref{4L-LL}(b) describes an evolution of Landau
level energies with increasing $\Delta$,
where we observe the triply-degenerate Landau levels 
of $p_+$ pockets with wide energy spacing, similarly to trilayer graphene. 

We can expand the Hamiltonian with respect to 
the center of each Dirac pocket,
and obtain the effective $2\times 2$ Hamiltonian
in the massive Dirac form of Eq.\ (\ref{eq_massive_dirac}).
The masses and the velocities are
shown in Fig.\ref{4L-LL}(c) and (d), respectively,
where we set the basis so that the chirality becomes $+1$ 
(i.e., $v_xv_y$ positive).
The masses at $p_+$ and $p_-$ are opposite in sign, 
so that the contributions to the Hall conductivity 
of Dirac cones cancel out in summation over a single valley.
Therefore, the single valley system
is a trivial insulator with zero Hall conductivity
in contrast to trilayer graphene.
Accordingly, we observe no Landau level crossing 
at charge neutral point in Fig.\ref{4L-LL}(b),
and if we look at the edge states in Fig.\ref{4L-LL}(e), 
there are no edge channel crossing in the bulk gap,
nor the counter-propagating flows of valley pseudo-spin.

\section{General odd-layer graphenes}
\label{sec_odd}

The approximate chiral symmetry in the presence of 
the external electric field argued in trilayer actually holds in 
any odd-layer Bernal multilayer graphenes.
As we see in the following,
if we only consider $\gamma_0,\gamma_1,\gamma_3,\Delta$ terms ,
the Hamiltonian of odd layered graphene 
is chiral symmetric with a nonzero chiral index
in the presence of magnetic field,
which means that the band is gapless in limit of zero magnetic field.

The Hamiltonian for Bernal stacked $N$-layer graphene
is decomposed into block diagonal form with 
effective monolayer and bilayer blocks with 
a unitary transformation \cite{koshino-ando07,koshino-ando08,koshino2009electronic,koshino-LL-2011}.
Let us define
\begin{eqnarray}
f_m(j)=c_m\sqrt{\frac{2}{N+1}}[1-(-1)^j]\sin \kappa_m j,
\nonumber\\
g_m(j)=c_m\sqrt{\frac{2}{N+1}}[1+(-1)^j]\sin \kappa_m j,
\label{generalN-bases}
\end{eqnarray}
where 
\begin{eqnarray}
\kappa_m=\frac{\pi}{2}-\frac{m\pi}{2(N+1)},
\nonumber \\
c_m=
\begin{cases}
1/2 & (m=0), \\
1/\sqrt 2 & (m \neq 0), \\
\end{cases}
\end{eqnarray}
 $j=1,2,...,N$ is the layer index,
$m$ is the block index given as 
$$
m=
\begin{cases}
1,3,5,...,N-1; & N= \mbox{even}, \\
0,2,4,...,N; & N = \mbox{odd}. \\
\end{cases}
$$
Then we take a basis 
\begin{eqnarray}
|\phi_m^{(X,{\rm odd})} \rangle=\sum_{j=1}^N f_m(j)|X_j \rangle,
\nonumber\\
|\phi_m^{(X,{\rm even})} \rangle=\sum_{j=1}^N g_m(j)|X_j \rangle,
\label{multi-layer-bases}
\end{eqnarray}
where $X=A$ or $B$. A superscript $(X,{\rm odd/even})$ 
indicates that the wavefunction is nonzero only on
the sublattice $X$ on the layer $j = $ odd/even.

With the basis set above, the Hamiltonian 
with $\Delta = 0$ is block diagonalized with $m$.
A block labeled by $m$ spanned by 
$
\{|\phi_m^{(A,{\rm odd})} \rangle,
|\phi_m^{(B,{\rm odd})} \rangle,
|\phi_m^{(A,{\rm even})} \rangle,
|\phi_m^{(B,{\rm even})} \rangle\}
$ 
is dictated as $H(\lambda_m)$ with $\lambda_m=2\cos\kappa_m$ and
\begin{equation}
H(\lambda)=
\begin{pmatrix}
0 & v \pi^\dagger & 0 & \lambda v_3 \pi \\
v \pi & 0 & \lambda \gamma_1 & 0 \\
0 & \lambda \gamma_1 & 0 & v \pi^\dagger \\
\lambda v_3 \pi^\dagger & 0 & v\pi& 0 \\
\end{pmatrix},
\label{blockH}
\end{equation}
where we only left the terms with $\gamma_0,\gamma_1,\gamma_3$,
For the case of $m=0$, 
only two bases $
|\phi_m^{(A,{\rm odd})} \rangle,
|\phi_m^{(B,{\rm odd})} \rangle
$  
survive due to $g_0(j)=0$,
and the corresponding block Hamiltonian is the first 2 by 2 component of Eq.(\ref{blockH}).

Now let us consider an odd-layer graphene with
the interlayer potential asymmetry $U(\Delta)$, 
i.e., the second term of Hamiltonian, Eq.\ (\ref{BernalH}).
When $N={\rm odd}$, we can easily show that
the basis function belonging to the block $m$
is either symmetric or antisymmetric
with respect to reflection in the central layer $j=(N+1)/2$,
and the parity is given by $(-1)^{(N-m-1)/2}$. \cite{koshino-LL-2011}
If we write down $U(\Delta)$ in this basis, 
the matrix element $\langle\phi_{m'}^{(X',p')} |U|\phi_m^{(X,p)} \rangle$
($p,p' =$ even or odd)
becomes non-zero only when $m'=m+4l+2(l:{\rm integer})$,
$X=X'$, and $p=p'$,
because $U$ is an odd function in the reflection,
and also diagonal in the original site representation.

Therefore, if we separate the basis functions 
into two groups $\circ,\bullet$ as:
\begin{eqnarray}
&&
(
|\phi_m^{(A,{\rm odd})} \rangle ,
|\phi_m^{(B,{\rm odd})} \rangle,
|\phi_m^{(A,{\rm even})} \rangle,
|\phi_m^{(B,{\rm even})} \rangle
) 
\nonumber\\
&&
\qquad\qquad
\in 
\begin{cases}
(\circ,\bullet,\circ,\bullet) \quad (m=4l)\\
(\bullet,\circ,\bullet,\circ) \quad (m=4l+2),
\end{cases}
\label{chiral-bases}
\end{eqnarray} 
then the Hamiltonian including only $\gamma_0,\gamma_1,\gamma_3,\Delta$
has matrix elements only between $\circ$ and $\bullet$, 
and thus is chiral symmetric.
Note that the chiral symmetry is not respected for even-layer graphenes
since the basis function labeled by $m$ cannot be categorized to either
even or odd parity, and 
the interlayer potential term gives rise to matrix elements 
connecting blocks $m$ and $m+4l$.

In the presence of magnetic field, the chiral index, i.e., 
the difference between the number of the zero-modes belonging to 
$\circ$ $(n_+)$  and those to $\bullet$ $(n_-)$,
can be easily obtained by considering the Landau level spectrum
with $\gamma_1$, $v_3$ and $\Delta$ all switched off,
since the chiral index never changes in such a continuous transformation.
The bilayer-like Hamiltonian block, Eq.\ (\ref{blockH}), 
is then consists of two monolayer-type $2\times 2$ diagonal blocks,
giving two zero-energy Landau levels
localized at the second and fourth (first and third) elements
in the valley $\xi = +(-)$.
Considering the base grouping in Eq.\ (\ref{chiral-bases}),
the difference in zero energy states in block $m$ 
is found to be $n_+-n_- = \mp 2 \xi \Phi$ 
for $m = 4l$ and $m = 4l+2$, respectively,
where $\Phi = eBS/h$ 
is the magnetic flux penetrating the system.
The monolayer-type block $m=0$ lacks the third and fourth elements,
giving $n_+-n_- = - \xi \Phi$.
As a result, the total chiral index in odd-layer graphene 
for valley $K_\xi$ is finally given by
\begin{eqnarray}
n_+-n_- &=& (-1 +2 -2 +2 \cdots)\xi \Phi
\nonumber\\
&=&
\begin{cases}
-\xi \Phi; & N=4l+1 \\
+\xi \Phi; & N=4l+3. \\
\end{cases}
\label{eq_index_general}
\end{eqnarray}

This states that at least one Landau level
remains at zero energy,
and thus in zero magnetic field,
the conduction and valence band touch at one $k$-point at least.
The sum of Berry phases in a single valley
coincides with $-\pi$ times the chiral index in units of $\Phi$,
that is $\pm \xi \pi$ for $N=4l+1$ and $4l+3$, respectively.
When the Dirac cones are gapped by including the additional band
parameters, the Hall conductivity per single valley must be non-zero,
because the number of Dirac cones per valley must be odd
to achieve total Berry phase $\pm \pi$.

The result might seem to contradict with 
the well-known fact that the Berry phase is $N\pi$ in 
$N$-layer graphene at $\Delta = 0$.
As argued, the Hamiltonian with $\Delta = 0$ is block diagonalized 
into independent monolayer-like and bilayer-like subsystems.
In each block, there is an ambiguity in the choice 
of bases for $\circ$ or $\bullet$, 
and the Berry phase of the block actually changes its sign
when $\circ$ and $\bullet$ are interchanged.
If we simply assign 
all $A$ and $B$ sublattices to $\circ$ and 
$\bullet$, respectively,  we obtain
$n_+ - n_- = (-1-2-2-2\cdots)\xi \Phi = -\xi N \Phi$
instead of Eq.\ (\ref{eq_index_general}).
In the presence of nonzero $\Delta$, on the other hand, 
the subsystems are mixed with each other
and then the grouping of Eq.\ (\ref{chiral-bases}) is the only possible
way to make the Hamiltonian chiral symmetric.



\section{Conclusion}
\label{sec_concl}

In Bernal multilayer graphene more than three layers,
an interplay of the gate electric field and the trigonal warping effect
gives rise to emergent Dirac cones in the low energy bands,
whose band velocity and Lifshitz transition energy are 
tunable by the gate voltage.
In trilayer graphene, in particular,
the low-energy effective theory shows that 
the valley Hall state is realized at the charge neutral point,
where single valley Hall conductivity is quantized at a non-zero half integer.
We have investigated the edge states at the zigzag interface, 
and demonstrated that the number of edge modes is
closely related to the bulk single valley Hall conductivity.
In four-layer graphene, gate-induced Dirac cones also appear,
though the system is a trivial insulator with zero 
valley Hall conductivity.
The non-trivial valley Hall state is generally found 
in odd-layer graphenes,
where the approximate chiral symmetry is 
responsible for the emergence of non-zero valley Hall 
conductivity.

The gate-induced Dirac cones should be experimentally accessible
directly by observing Landau levels with wide energy spacing.
\cite{sadowski2006landau,jiang2007infrared,morimoto-bilayer-12}
Also, the single valley Hall conductivity 
argued here is expected to be observable 
in the transport through the edge modes at a zigzag interface,
while a valley mixing caused by a concentration of atomic-scale
scatterers or a presence of armchair edge would wash out the effect.
There the conductance is related to the number of edge channels,
and the topological transition at $\Delta_c$ should be
observed as a change in the conductance.
Since the helical edge modes appearing in 
the gated multilayer graphene carry valley pseudospins,
modulation of edge modes through the gate voltage
could be a way to electrically control the valley polarized transport.

\section*{Acknowledgments}

Authors thank helpful discussions with Akira Furusaki.
This work was supported by Grants-in-Aid for Scientific Research,
No.24840047 (TM), No.24740193 (MK)  from JSPS.
 
\appendix
\section{Index theorem for odd-layer graphenes}\label{sec:chiral}

Here we show that the chiral index of general odd-layer graphenes
can be written in terms of the total gauge flux penetrating the system,
in a similar way to the argument for usual Dirac Hamiltonian.
\cite{jackiw-rebbi-index77,fujikawa79,fujikawa-anomaly}
The chiral symmetric Hamiltonian 
of odd-layer graphene with interlayer asymmetric potential 
is written as
\begin{eqnarray}
H=
\begin{pmatrix}
0 & D_- \\
D_+ & 0
\end{pmatrix}
,
\end{eqnarray}
where $D_+ = (D_-)^\dagger$ is a $N\times N$ matrix.
The chiral symmetry is then expressed by
$\Gamma H + H \Gamma=0$
with the chiral operator $\Gamma$,
$$
\Gamma=\begin{pmatrix}
\1_N&0\\
0&-\1_N\\
\end{pmatrix},
$$
where $\1_N$ is $N\times N$ unit matrix.
We can take zero modes of $H$ as an eigenvector $\psi_\pm$ of
$\Gamma$ with eigenvalue $\pm 1$, respectively.
If we write the number of chiral zero modes $\psi_\pm$ as $n_\pm$,
the chiral index $\nu$ is defined as
$$
\nu=n_+ - n_- = \mbox{Tr} \Gamma f(H^2/M^2),
$$
where $f$ is a regularization function, which is 
smooth and monotonically decreasing function with $f(0)=1$ and $f(\infty)=0$, 
and $M$ is an ultraviolet cutoff. 
The action of $f$ to the matrix is 
defined through its action onto the eigenvalues, 
as seen if we take $f(z)=e^{-z}$.

If we take plane waves $\exp(i \Vec{k} \cdot \Vec{x})$ as a basis for the 
spatial direction, we have 
\begin{eqnarray*}
\nu&=&
\mbox{Tr} \Gamma f(H^2/M^2) 
\\
&=& \mbox{Tr} [f(D_-D_+/M^2)-f(D_+D_-/M^2)] \\
&=& 
\int d^2 x
\int \frac{d^2 k}{(2\pi)^2}
\mbox{tr} \,
e^{-i k x}
[f(D_-D_+/M^2)\nonumber\\
&& \qquad \qquad \qquad -f(D_+D_-/M^2)] e^{i k x}
\label{index-right}
\end{eqnarray*}
where Tr means a trace over all the states 
while tr is a trace over the
layer and site indeces or, equivalently, 
$m$, $X$ and even/odd indices in Eq.(\ref{multi-layer-bases}).

The operator $\GVec{\pi}$ acts on a plane wave 
to give
$e^{-ikx} \GVec{\pi}e^{ikx}  = \hbar \Vec{k}+ \GVec{\pi}$.
Since $D_\pm$ contains at most first order derivative 
terms, we can write
\begin{eqnarray*}
&& e^{-i k x} D_+ e^{i k x}=D_0  |\Vec{k}| + D_+ \\
&& e^{-i k x} D_- e^{i k x}=D_0^\dagger  |\Vec{k}| + D_-,
\end{eqnarray*}
where $D_0$ is a c-number matrix and of zero-th order in $|\Vec{k}|$, 
but may depend on a polar angle of $\Vec{k}$.
Since $D_0$ is a square matrix, 
it can be written in a singular value decomposition 
$$
D_0=U \Sigma V^\dagger
$$ with unitary matrices $U,V$ and 
a diagonal matrix $\Sigma=\mbox{diag}(\sqrt{\lambda_i})$ $(\lambda_i \ge 0)$.

The action of $D_- D_+$ on a plane wave is then described as
$$
e^{-i k x} D_\mp D_\pm e^{i k x}= G_\pm k ^2 + F_\pm,
$$
where 
\begin{eqnarray*}
&&G_+=D_0^\dagger D_0=V \Sigma^2 V^\dagger \\
&&G_-=D_0 D_0^\dagger=U \Sigma^2 U^\dagger, \\
\end{eqnarray*}
are c-number matrices of zero-th order in $|\Vec{k}|$.
$F_\pm$ are matrices including the operator $\GVec{\pi}$,
and up to first order of $|\Vec{k}|$.

Having in mind that the ultraviolet behavior ($\hbar v k \sim M$) is important
for the contribution of $D_-D_+$ term to the chiral index, 
we obtain
\begin{eqnarray*}
&& \mbox{tr} \{ e^{-i k x} f(D_- D_+/M^2) e^{i k x} \} \\
&=& \mbox{tr} \{ V f(k^2 \Sigma^2/M^2+ V^\dagger F_+ V/M^2) V^\dagger \} \\
&=& \mbox{tr} \{f( k^2 \Sigma^2/M^2)+f'( k^2 \Sigma^2/M^2)(V^\dagger F_+ V)/M^2 \} 
\\ && \hspace{18em}
 +O(M^{-4}) \\
&=& \sum_i\{f( k^2 \lambda_i/M^2)+f'( k^2 \lambda_i/M^2)(V^\dagger F_+ V)_{ii}/M^2)\}
\\ && \hspace{18em}
 +O(M^{-4})
\end{eqnarray*}
With a similar form for $D_+D_-$ term, the chiral index with $M \to \infty$ reduces to
\begin{eqnarray}
\nu
&=&  
\int d^2 x
\int \frac{d^2 k}{(2\pi)^2} \sum_i f'(k^2
\lambda_i/M^2)\nonumber\\
&& \hspace{15mm}\times (V^\dagger F_+ V - U^\dagger F_- U)_{ii}/M^2. 
\label{index-gauge}
\end{eqnarray}


As argued in Sec.\ \ref{sec_odd}, the Hamiltonian 
of multilayer graphene with $\Delta = 0$
is decomposed into block diagonal form with 
effective monolayer and bilayer blocks labeled by $m$,
and when the number of layers is odd,
$\Delta$ always enters in the off-diagonal blocks.
Then $G_\pm$ is block-diagonal because
it is independent of $\Delta$, and so $U$ and $V$ are.
Therefore, although block off-diagonal terms in $F_\pm$ arise from $\Delta$,
they do not contribute to the sum of $(V^\dagger F_+ V-U^\dagger F_- U)_{ii}$,
and thus we only  have to compute the chiral index for each block separately and add them up to obtain the overall chiral index.

{\it Monolayer block $(m=0)$.} Using the commutation relation $[\pi,\pi^\dagger]=-\xi2\hbar eB$,
\begin{eqnarray*}
D_- D_+/v^2 &=& \pi^\dagger \pi = \Vec \pi^2 + \xi \hbar e B \\
D_+ D_-/v^2 &=& \pi \pi^\dagger = \Vec \pi^2 - \xi \hbar e B \\
G_\pm &=& \hbar^2 v^2 \\
F_\pm &=& v^2(\GVec{\pi}^2 + 2\hbar \Vec{k}\cdot\GVec{\pi} ) 
\pm \xi v^2 \hbar e B. \\
\end{eqnarray*}
From Eq.(\ref{index-gauge}),
\begin{eqnarray*}
\nu
&=& \int d^2x \int \frac{d^2 k}{(2\pi)^2} f'(\hbar^2 v^2 k^2/M^2)(2\xi v^2 \hbar
eB)/M^2 \\
&=& - \xi \Phi,  
\end{eqnarray*}
where $\Phi = \int d^2x\, eB/h$ is the magnetic flux penetrating 
the system.
When the magnetic field is uniform, 
this is the Landau level degeneracy of $n=0$ Landau level,
and its sign reflects that the level
is assigned to $\psi_- (\psi_+)$ at $K^+(K^-)$ valley.

{\it Bilayer block $(m\neq0)$.}
For simplicity, we set $v=1$ and $\lambda=1$,
and compensate it by redefining $M/v \to M$,$\lambda v_3/v \to v_3$ and $\lambda \gamma_1/v \to \gamma_1$.
If we rewrite the block Hamiltonian Eq. (\ref{blockH}) 
in an order of bases 
$
|\phi_m^{(A,{\rm odd})} \rangle ,
|\phi_m^{(A,{\rm even})} \rangle,
|\phi_m^{(B,{\rm even})} \rangle,
|\phi_m^{(B,{\rm odd})} \rangle
$ as
\begin{eqnarray*}
H=
\begin{pmatrix}
0 & D_- \\
D_+ & 0\\
\end{pmatrix} 
=
\begin{pmatrix}
0 & 0 & v_3 \pi & \pi^\dagger \\
0 & 0 & \pi^\dagger & \gamma_1 \\
v_3 \pi^\dagger & \pi & 0 & 0 \\
\pi & \gamma_1 & 0 & 0 \\
\end{pmatrix} ,
\end{eqnarray*}
we have 
\begin{eqnarray*}
\hspace{-1em}
D_- D_+ &=&
\begin{pmatrix}
\Vec \pi^2 (1+v_3^2) & \gamma_1 \pi^\dagger + v_3 \pi ^2 \\
\gamma_1 \pi + v_3 (\pi^\dagger) ^2 & \Vec \pi^2 + \gamma_1^2
\end{pmatrix}
+
\xi \hbar e B
\begin{pmatrix}
1-v_3^2 & 0\\
0 & 1 
\end{pmatrix} 
\\
\hspace{-1em}
D_+ D_- &=&
\begin{pmatrix}
\Vec \pi^2 (1+v_3^2) & \gamma_1 \pi + v_3 (\pi^\dagger) ^2 \\
\gamma_1 \pi^\dagger + v_3 \pi ^2 & \Vec \pi^2 + \gamma_1^2
\end{pmatrix}
- \xi \hbar e B
\begin{pmatrix}
1-v_3^2 & 0\\
0 & 1 
\end{pmatrix} .
\end{eqnarray*}
The action on the plane wave basis is then given by
\begin{eqnarray*}
e^{-i k x} D_- D_+ e^{i k x} &=& G_+ k^2 + \tilde{F} + F \\ 
e^{-i k x} D_+ D_- e^{i k x} &=& (G_+ k^2 + \tilde{F})^\mathrm{T} - F \\ 
G_+ &=&
\hbar^2 \begin{pmatrix}
1+v_3^2 & v_3 e^{2 i \t} \\
v_3 e^{- 2 i \t} & 1 \\
\end{pmatrix}, \\
F &=& \xi \hbar e B
\begin{pmatrix}
1-v_3^2 & 0\\
0 & 1 
\end{pmatrix}, 
\end{eqnarray*}
where $\tilde{F}$ is a matrix including $\GVec{\pi}$,
of which expression (not presented) is not important 
in the following argument.

With a relation $\mbox{tr} f(A)=\mbox{tr} f(A^{\mathrm{T}})$, an explicit calculation of Eq.(\ref{index-right}) shows that 
\begin{eqnarray*}
\nu
&=&  \int d^2 x
\int \frac{d^2 k}{(2\pi)^2} \sum_i f'(k^2 \lambda_i/M^2)
\\ && \times
[V^\dagger (\tilde F +F) V - (V^{\mathrm{T}})^\dagger (\tilde F^{\mathrm{T}} - F) V^{\mathrm{T}}]_{ii}/M^2 \\
&=&  \int d^2 x \int \frac{d^2 k}{(2\pi)^2} \sum_i f'(k^2 \lambda_i/M^2)
(2V^\dagger F V )_{ii}/M^2 \\
&=&  - \frac{1}{2\pi} \int d^2 x  \sum_i \frac{1}{\lambda_i} 
(V^\dagger F V )_{ii} \\
&=&  - \frac{1}{2\pi}\int d^2 x \, {\rm tr}\,(G_+^{-1}F)
= -2\xi\Phi.
\end{eqnarray*}
Thus, the chiral index of bilayer block is twice of
that of monolayer block,
and inclusion of $v_3$ does not affect the result.
This is naturally expected because the chiral index 
is a topological number and never changes in a continuous deformation.

If we combine these results for monolayer block and bilayer blocks
noting that the orders of the chiral bases 
for each block (Eq.(\ref{chiral-bases})),
we find that the chiral index for biased $N$ (odd) layered graphene 
is given by 
\begin{eqnarray*}
n_+-n_- &=& (-1 +2 -2 +2 ....)\xi \Phi
\\
&=&
\begin{cases}
-\xi \Phi; & N=4l+1 \\
+\xi \Phi; & N=4l+3. \\
\end{cases}
\end{eqnarray*}
This exactly coincides with Eq.\ (\ref{eq_index_general}),
while the present argument is more general and valid 
for non-uniform magnetic field $B(x,y)$.


\section{Derivation of the edge modes from the effective mass Hamiltonian}\label{sec:edge}

When the Hamiltonian is linear in $\Vec{k}$,
it is possible to obtain the edge state energies in the bulk gap, 
by searching for the evanescent modes satisfying a boundary condition
at the interface.
We consider a $2N \times 2N$ Hamiltonian matrix
$H(\hat{k}_x,\hat{k}_y)$
with $\hat{\Vec{k}} = -i\nabla$,
and assume it is linear in $\hat{\Vec{k}}$.
It is expressed as
\begin{equation}
 H = A \hat{k}_y  + B(\hat{k}_x),
\end{equation}
where $A$ and $B$ is $2N \times 2N$ matrices
and $A$ is independent of $\hat{k}_x$.
We assume the system is periodic  in $x$
and replace $\hat{k}_x$ with its eigenvalue $k_x$.
$H$ is regarded as one-dimensional Hamiltonian
with a parameter $k_x$.
The Schr\"{o}dinger equation, $H\Vec{F}(y) = \vare\Vec{F}(y)$
is transformed to
\begin{equation}
 \frac{\partial}{\partial y} \Vec{F}(y) = 
iA^{-1}[-B(k_x)+\vare]\Vec{F}(y) \equiv M(k_x,\vare) \Vec{F}(y),
\end{equation}
with $M = iA^{-1}[-B(k_x)+\vare]$.
Let $\kappa^{(\alpha)}$ and 
$\Vec{f}^{(\alpha)} (\alpha = 1,\cdots 2N)$ 
the eigenvalues and right eigenvectors of 
the matrix $M(k_x,\vare)$, 
The corresponding wave function becomes
\begin{equation}
 \Vec{F}^{(\alpha)}(y) = \exp(\kappa^{(\alpha)} y) \Vec{f}^{(\alpha)}.
\end{equation}
Generally $\kappa^{(\alpha)}$ is a complex number,
and the state is a bulk mode when $\kappa^{(\alpha)}$
is pure imaginary, and an evanescent mode otherwise.
When the bulk spectrum of $H(k_x,k_y)$ is fully gapped 
at particular $k_x$, and $\vare$ is inside the gap,
we have $2N$ evanescent modes with
$N$ modes of ${\rm Re}\, \kappa^{(\alpha)} > 0$
and another $N$ modes of ${\rm Re}\, \kappa^{(\alpha)} < 0$.
When we consider the half-infinite system in the region $y\geq 0$,
an edge state localized near $y=0$, if it exists,
should be written as a linear combination of the
states with ${\rm Re} \, \kappa^{(\alpha)} < 0$. 
When the indeces $\alpha = 1,2,\cdots,N$ are assigned to the 
modes of ${\rm Re} \, \kappa^{(\alpha)} < 0$,
it is written as
\begin{equation}
 \Vec{F}(y) = \sum_{\alpha=1}^{N}
C^{(\alpha)} \exp(\kappa^{(\alpha)} y) \Vec{f}^{(\alpha)}.
\label{eq_F}
\end{equation}
The boundary condition at $y=0$ for the wavefunction $\Vec{F}$
is composed of $N$ linear equations with respect to
$2N$-dimensional vector $\Vec{F}(0)$.
This is written as $ D \Vec{F}(0) = 0$,
with $D$ being a $N\times 2N$ constant matrix.
For the trilayer graphene with zigzag edge, for example,
the boundary condition at $y=0$ is 
$F_1(0)= F_3(0)= F_5(0)=0$,
so that $D$ becomes
\begin{equation}
 D = 
\begin{pmatrix}
1 & 0 & 0 & 0 & 0 & 0 \\
0 & 0 & 1 & 0 & 0 & 0 \\
0 & 0 & 0 & 0 & 1 & 0 
\end{pmatrix}.
\end{equation}

By using Eq.\ (\ref{eq_F}),
the boundary condition is written as
\begin{equation}
X
\begin{pmatrix}
 C^{(1)}\\
 C^{(2)}\\
\vdots\\
 C^{(N)}
\end{pmatrix}
=
\begin{pmatrix}
0\\
0\\
\vdots\\
0
\end{pmatrix}
\end{equation}
where $X = X(k_x, \vare)$ is a $N \times N$ matrix defined by
\begin{equation}
X =  D\left( 
\Vec{f}^{(1)}, \Vec{f}^{(2)}, \cdots, \Vec{f}^{(N)}
\right) 
\end{equation}
The equation has a non-trivial solution when $\det X(k_x, \vare) =0$.
The edge mode energy can be found by
tracing $\det X(k_x, \vare)$ throughout the energy gap,
for each fixed $k_x$. 
When more than two edge modes are degenerate at the energy $\vare$,
the number of degeneracy is found by $N - {\rm Rank} X(k_x,\vare)$.



\bibliography{mybib12-7,gate,koshino}

\end{document}